\documentclass[11pt,a4paper]{article}
\usepackage[utf8]{inputenc}
\usepackage[margin=2.1cm]{geometry} % minimum requirement in 2cm, 2.1cm for margin!
\usepackage{graphicx} 
\usepackage{titling}
\usepackage{fancyhdr}
\usepackage{color}
\usepackage{wrapfig}
\usepackage{hyperref}
\usepackage{subfig}
\usepackage{setspace}
\usepackage{amssymb}
\usepackage[]{caption}
\newcommand\farcsec{\mbox{$.\mkern-4mu^{\prime\prime\mkern-2mu}$}}% 
\newcommand\arcmin{\mbox{$^\prime$}}% 
\newcommand\arcsec{\mbox{$^{\prime\prime}$}}% 

\preauthor{}
\postauthor{}
\author{} % No author
\date{}   % No date
\predate{}
\postdate{ 
  \begin{center}
  \vspace{-0.1in}
    \large\normalfont \textbf{Thematic Area}: ASTRO
  \end{center}
  \vspace{-2em}
}

\title{\textbf{Illuminating the Physics of Cosmic Origin and Evolution}\\ 
\Large \textbf{--- SIRMOS:} \textbf{S}atellite for \textbf{I}nfra\textbf{R}ed \textbf{M}ulti-\textbf{O}bject \textbf{S}pectroscopy}
\fancypagestyle{plain}{
    \fancyhf{} 
    \fancyfoot[L]{\thedate}
    \fancyhead[L]{UK Space Frontiers 2035 White Paper}
}

\makeatletter
\newcommand*{\rom}[1]{\expandafter\@slowromancap\romannumeral #1@}
\newcommand\ion[2]{#1$\;${%
\ifx\@currsize\normalsize\small \else
\ifx\@currsize\small\footnotesize \else
\ifx\@currsize\footnotesize\scriptsize \else
\ifx\@currsize\scriptsize\tiny \else
\ifx\@currsize\large\normalsize \else
\ifx\@currsize\Large\large
\fi\fi\fi\fi\fi\fi
\rmfamily\rom{#2}}\relax}% 
\makeatother

\begin{document}

\maketitle
\thispagestyle{plain}

\begin{figure}[htbp]
    \centering
    \includegraphics[width=0.55\textwidth]{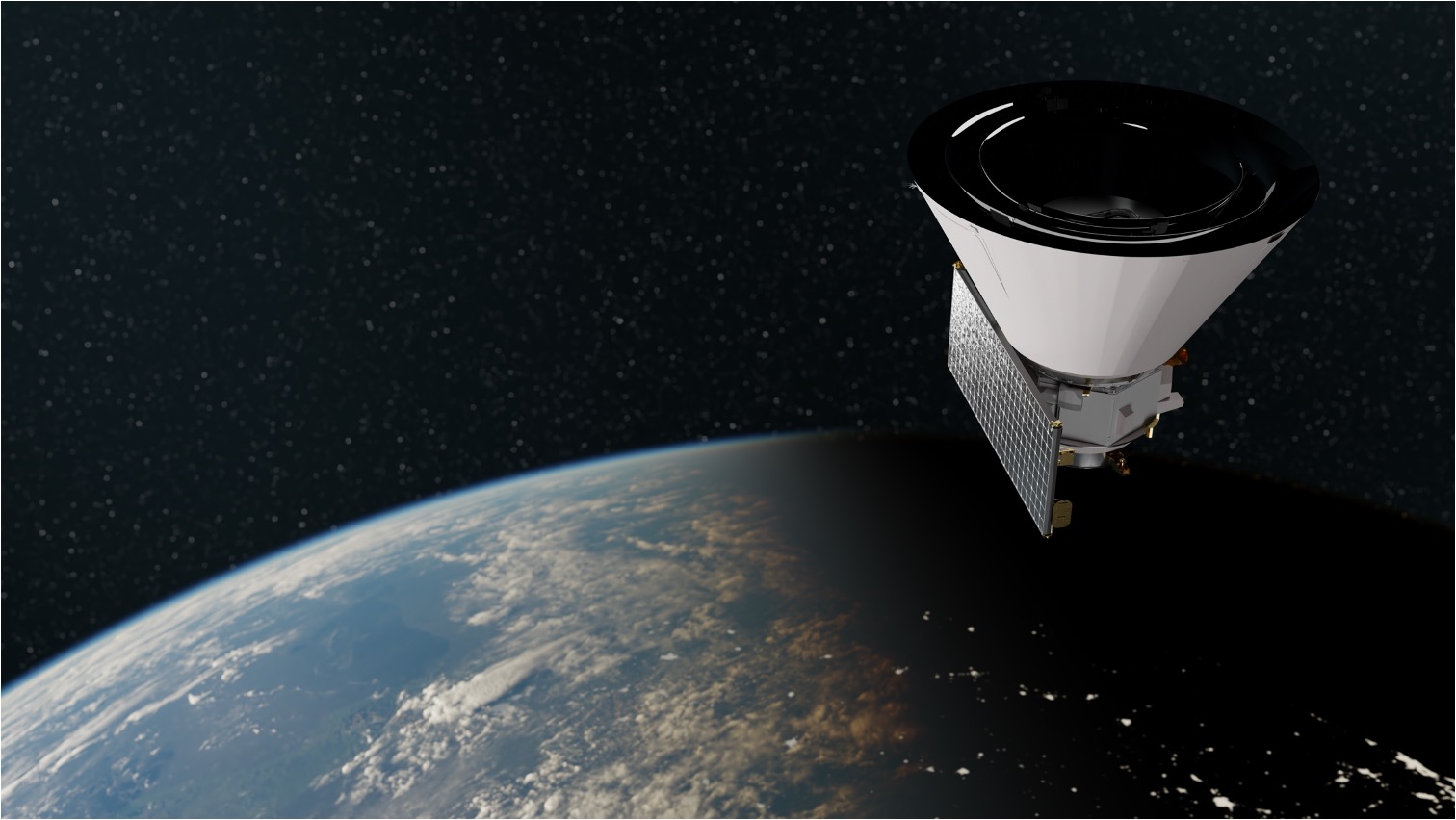}
\end{figure}

\vspace{-0.05in}
\noindent
{\footnotesize 
    \textbf{UK authors:} 
    Florian Beutler (Edinburgh),
    Eva-Maria Mueller\footnote{em675@sussex.ac.uk} (Sussex),
    Seshadri Nadathur (Portsmouth), 
    David Alonso (Oxford),
    Tessa Baker (Portsmouth), 
    Sownak Bose (Durham), 
    Rebecca Canning (Portsmouth),
    Shaun Cole (Durham), 
    Fergus Cullen (Edinburgh),
    Willem Elbers (Durham),
    Pedro Ferreira (Oxford), 
    Carlos Frenk (Durham), 
    Oscar Gonzalez (UKATC),
    Or Graur (Portsmouth),
    Boryana Hadzhiyska (Cambridge),
    Alex Hall (Edinburgh), 
    Catherine Heymans (Edinburgh),
    Sergey Koposov (Edinburgh),
    Kazuya Koyama (Portsmouth),
    Ofer Lahav (UCL), 
    Baojiu Li (Durham),
    Avery Meiksin (Edinburgh),
    Johannes Noller (UCL), 
    John Peacock (Edinburgh), 
    Alkistis Pourtsidou (Edinburgh),
    Giorgio Savini (UCL),
    Andy Taylor (Edinburgh),
    Rita Tojeiro (St Andrews), 
    David Wands (Portsmouth)\\
     % (more UK co-authors here)
    \noindent
    \textbf{International authors:} Yun Wang\footnote{Questions regarding the SIRMOS mission should be addressed to wang@ipac.caltech.edu} (Caltech/IPAC),
    Massimo Robberto (STScI), 
    Gregory Wirth (BAE Systems), 
    Mark Dickinson (NSF NOIRLab), 
    Thomas Greene (Caltech/IPAC),
    Jeffrey Kruk (GSFC), 
    Will Percival (Waterloo),
    Andreas Faisst (Caltech/IPAC), 
    Lynne Hillenbrand (Caltech),
    Jeyhan Kartaltepe (RIT),
    Nikhil Padmanabhan (Yale),
    Lado Samushia (Kansas State),
    Lee Armus (Caltech/IPAC), 
    Andrew Benson (Carnegie), 
    Micol Bolzonella (INAF Bologna), 
    Samuel Brieden (RWTH Aachen), 
    Jarle Brinchmann (ESO), 
    Robert Content (AAO), 
    Emanuele Daddi (CEA Saclay),
    Kyle Finner (Caltech/IPAC), 
    Andrew Hearin (ANL), 
    Cullan Howlett (Univ. of Queensland), 
    Jon Lawrence (AAO), 
    Gregory Mosby (GSFC), 
    Zoran Ninkov (RIT), 
    Ken Osato (Japan), 
    Casey Papovich (Texas A\&M), 
    Jack Piotrowski (Carnegie),   
    Lucia Pozzetti (INAF Bologna),
    Alvise Raccanelli (Padova), 
    Jason Rhodes (JPL), 
    Shun Saito (Missouri S\&T), 
    Hee-Jong Seo (Ohio Univ.), 
    Zachary Slepian (Univ. of Florida), 
    Steve Smee (JHU)
}

%\vspace{-0.2in}

\centerline{\bf Abstract} 
\begin{spacing}{0.9} 

{\small Understanding the Universe's origins and evolution remains one of the most fundamental challenges in modern cosmology. This white paper explores three key science priorities in this field: unravelling the physics of cosmic inflation, investigating the accelerating expansion of the Universe, and precisely measuring the sum of the neutrino masses. Achieving these goals requires a dedicated survey to map the large-scale structure at high redshift in unprecedented detail. We describe how this can be achieved through a mission concept called SIRMOS, providing a high-throughput, highly multiplexed spectroscopic capability to obtain accurate redshifts for over 100 million galaxies over a wide sky area. Such a survey would leverage the deepest existing wide-area photometric catalogues for targeting, with spectra offering continuous 1.25--2.5~$\mu$m wavelength coverage at moderate resolution, allowing precise redshift measurements in the $1<z<4$ range with minimal bias. We outline the scientific opportunities this presents. Recent years have seen significant advances in instrumentation, including digital micromirror devices, complex telescope mirrors, large detector arrays, and data processing pipelines. While these technologies have been demonstrated in terrestrial applications, such a survey is a unique opportunity to apply these proven capabilities in space to address fundamental questions in cosmology. Participation in such a mission will simultaneously deliver a compelling science case, help align UK Space Agency and STFC strategies, demonstrate the UK's growing capability in end-to-end space missions, and strengthen the national space economy through high-value industrial participation.}
\end{spacing}

\newpage

\section{Scientific motivation and objectives}
\vspace{-0.1in}

Our understanding of the evolution of the Universe has advanced enormously in the past few decades. We now believe it has undergone two distinct epochs of accelerated expansion. The first, inflation, occurred in the early Universe and set the initial conditions for the formation of all cosmic structure. The second, driven by dark energy, dominates the Universe today and will determine its ultimate fate. Both phenomena require the existence of energy components with negative pressure, a profound result which necessitates physics beyond our current Standard Model. Despite recent advances, the origins and properties of the fields responsible for these epochs remain unknown.

While cosmology has advanced primarily through measurement of the cosmic microwave background (CMB) and galaxy surveys covering relatively low redshifts, $z\lesssim2$, a wide-area survey mapping the high-redshift Universe at high sampling density through spectroscopic redshift measurements for $\sim100$ million galaxies over $1<z<4$ would allow us for the first time to precisely reconstruct cosmic evolution in this critical time period, illuminating both questions of cosmic origin and the fundamental physical laws operating in the Universe today.

\vspace{-0.1in}
\subsection{Science Goal 1:  Uncover how cosmic inflation gave rise to the large-scale structure of the Universe.}

\begin{figure}[hb!]
    \centering
    \vspace{-0.15in}
   \includegraphics[width=0.4\linewidth]{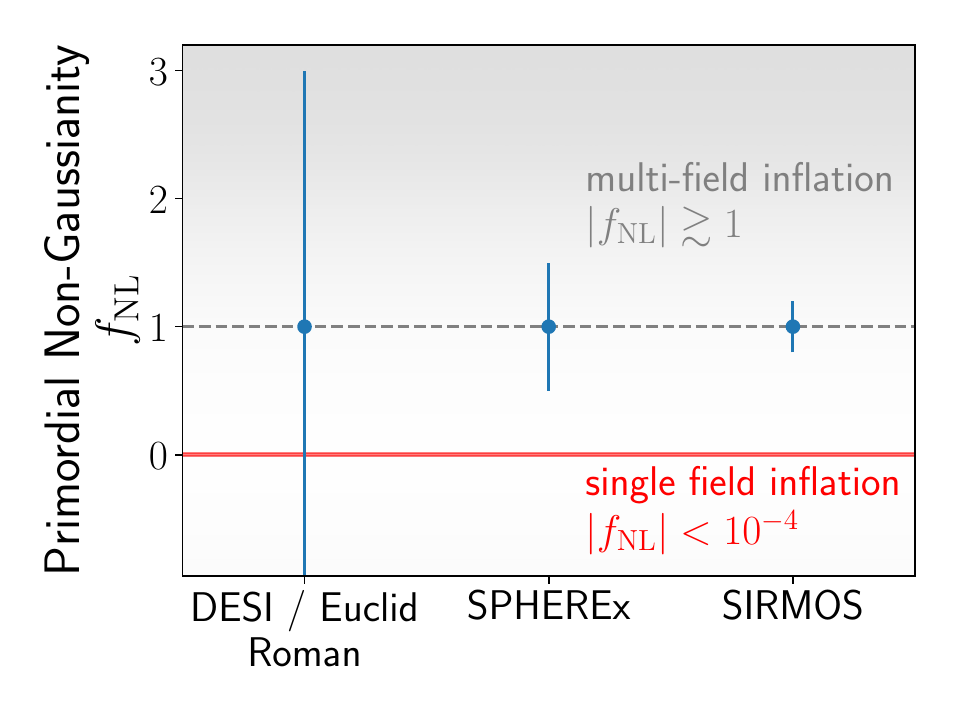}    \hspace{0.4in}
    \includegraphics[width=0.4\linewidth]{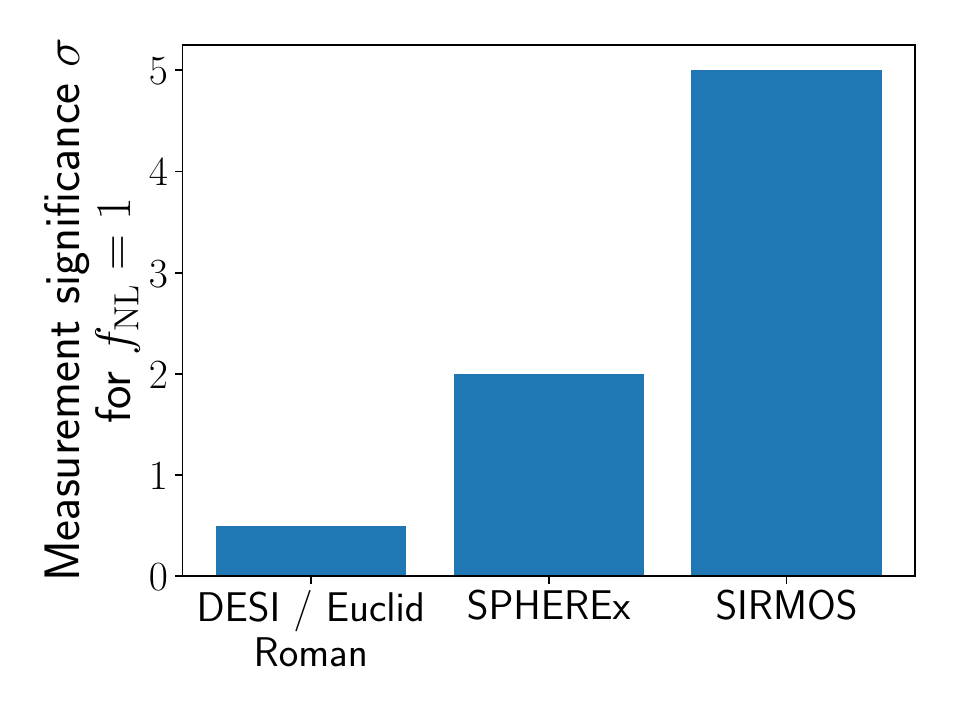}    
\vspace{-0.2in}
    \caption{Forecast constraints on non-Gaussianity parameter $f_\mathrm{NL}$ from various surveys (left), and the achievable measurement significance (right), assuming a target $f_\mathrm{NL}=1$. SIRMOS is designed to be able to detect $|f_{\mathrm{NL}}|=1$ at 5$\sigma$, thus distinguishing between single-field and multi-field inflation. 
    }
    \label{fig:placeholder}
    \vspace{-0.1in}
\end{figure} 

Inflation provides the leading explanation for the origin of cosmic structure, supported by observations of the near scale-invariance of the primordial power spectrum and the existence of super-horizon fluctuations in the CMB. Yet these successes do not reveal the details of the inflationary potential, nor even how many dynamical fields shaped the early Universe.

While the statistics of the fluctuations generated by inflation, which gave rise to cosmic structures, are well described by a Gaussian random field to first order at high precision, the possible existence of small non-Gaussianities provides a powerful discriminant distinguishing classes of inflationary models. The amplitude of these non-Gaussianities, quantified in terms of the local primordial non-Gaussianity (PNG) parameter, $f_\mathrm{NL}$, provides a means to distinguish profoundly different physical scenarios. Models with multiple fields in which the curvature perturbations are sourced by a field other than the inflaton generically predict $|f_\mathrm{NL}|\gtrsim1$, while single-field slow-roll inflation predicts $f_\mathrm{NL}\simeq0$ \cite{2014arXiv1412.4671A}. This sets a natural target for surveys.
\\
\underline{\bf Objective 1:} \textbf{Reach the sensitivity to detect $|f_\mathrm{NL}| = 1$  at $5\sigma$ significance.}
The best current measurements from the CMB have achieved $f_\mathrm{NL}=-0.9\pm5.1$ (68\% CL) \cite{2020A&A...641A...9P}, consistent with Gaussianity but insufficiently precise to rule out multi-field models of inflation. Substantial improvements in this precision can only come from accessing vastly more modes on ultra-large scales from galaxy clustering, specifically through redshift surveys covering exceptionally large cosmic volumes at high redshifts. Our forecasts show that a spectroscopic survey targeting a high density of observed galaxies over a wide sky area and covering $1<z<4$ will be able to achieve the sensitivity to detect $|f_\mathrm{NL}|=1$ at $5\sigma$ significance, exceeding the precision of other facilities (Figure 1) and \textbf{decisively determining whether inflation was driven by a single scalar field or more complex dynamics}. This drives the proposed design of SIRMOS (Section 2).

\vspace{-0.1in}
\subsection{Science Goal 2: Uncover the fundamental physics of dark energy. } 

\begin{figure}
    \centering
    \includegraphics[width=0.32\linewidth]{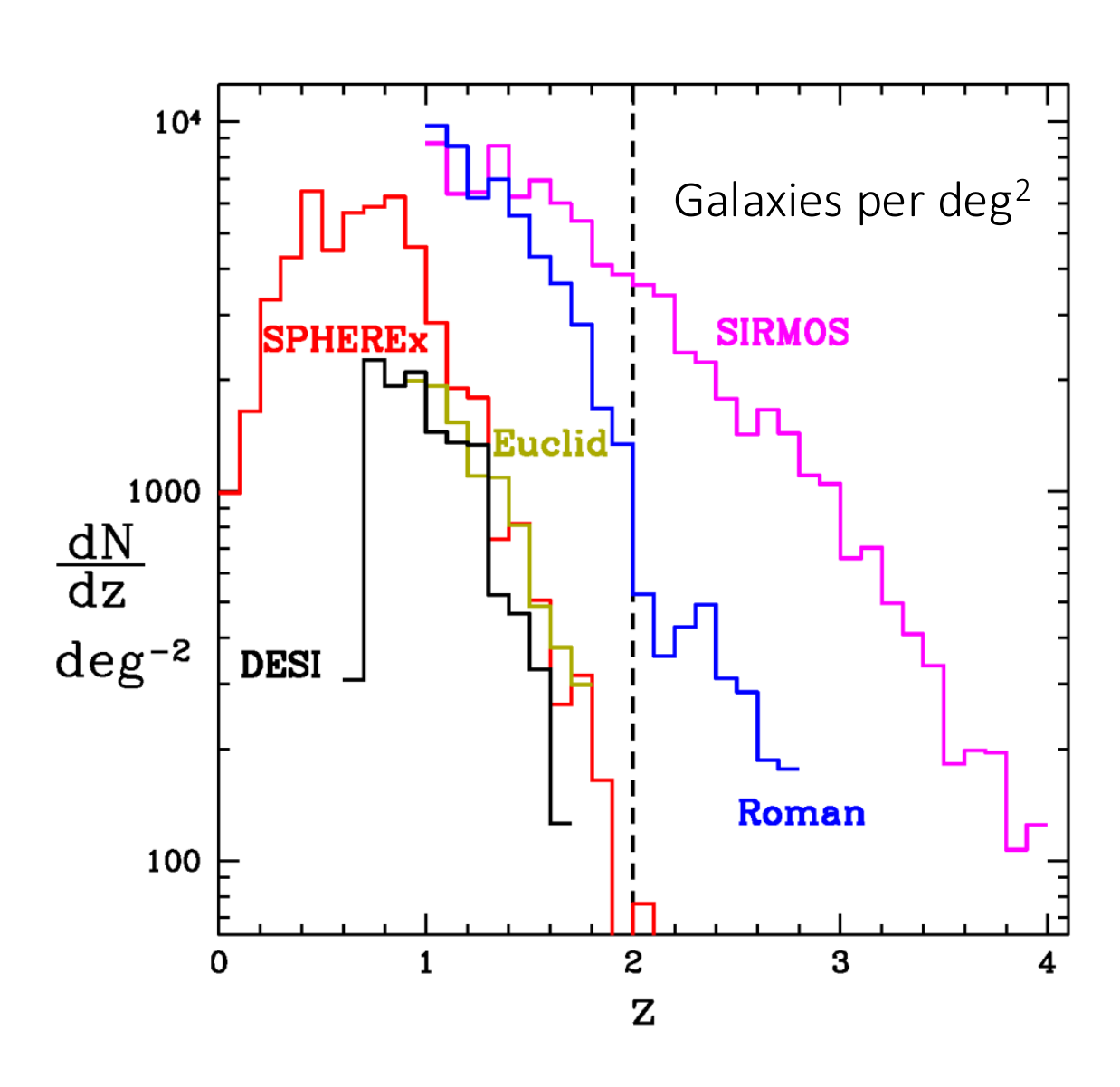}
    \includegraphics[width=0.32\linewidth]{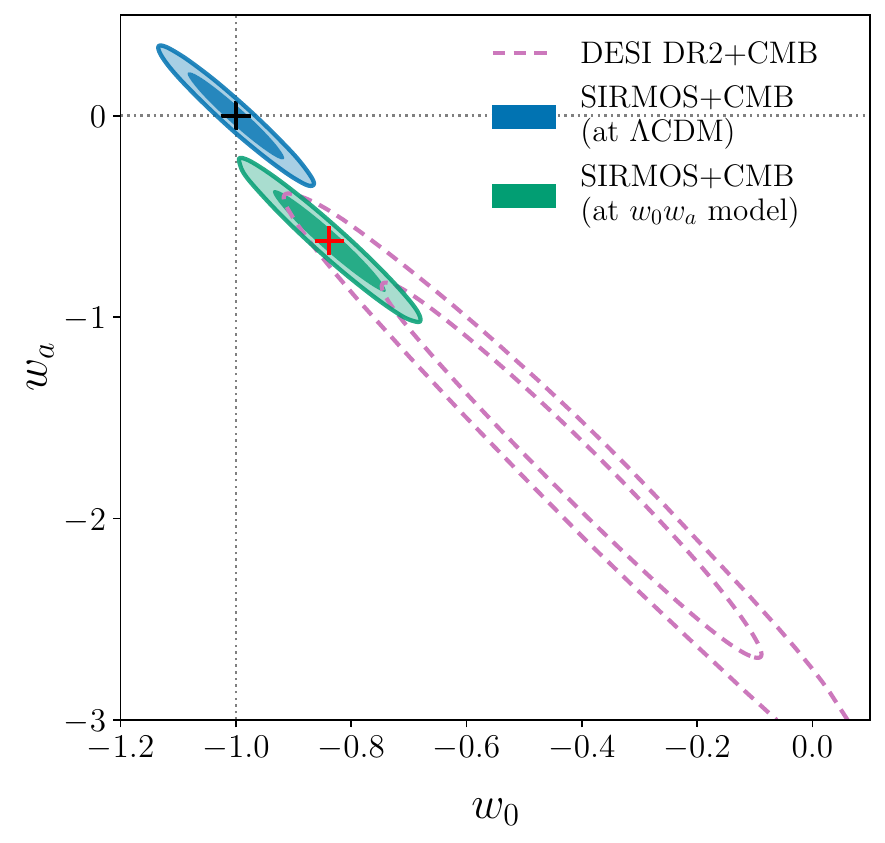}
    \includegraphics[width=0.32\linewidth]{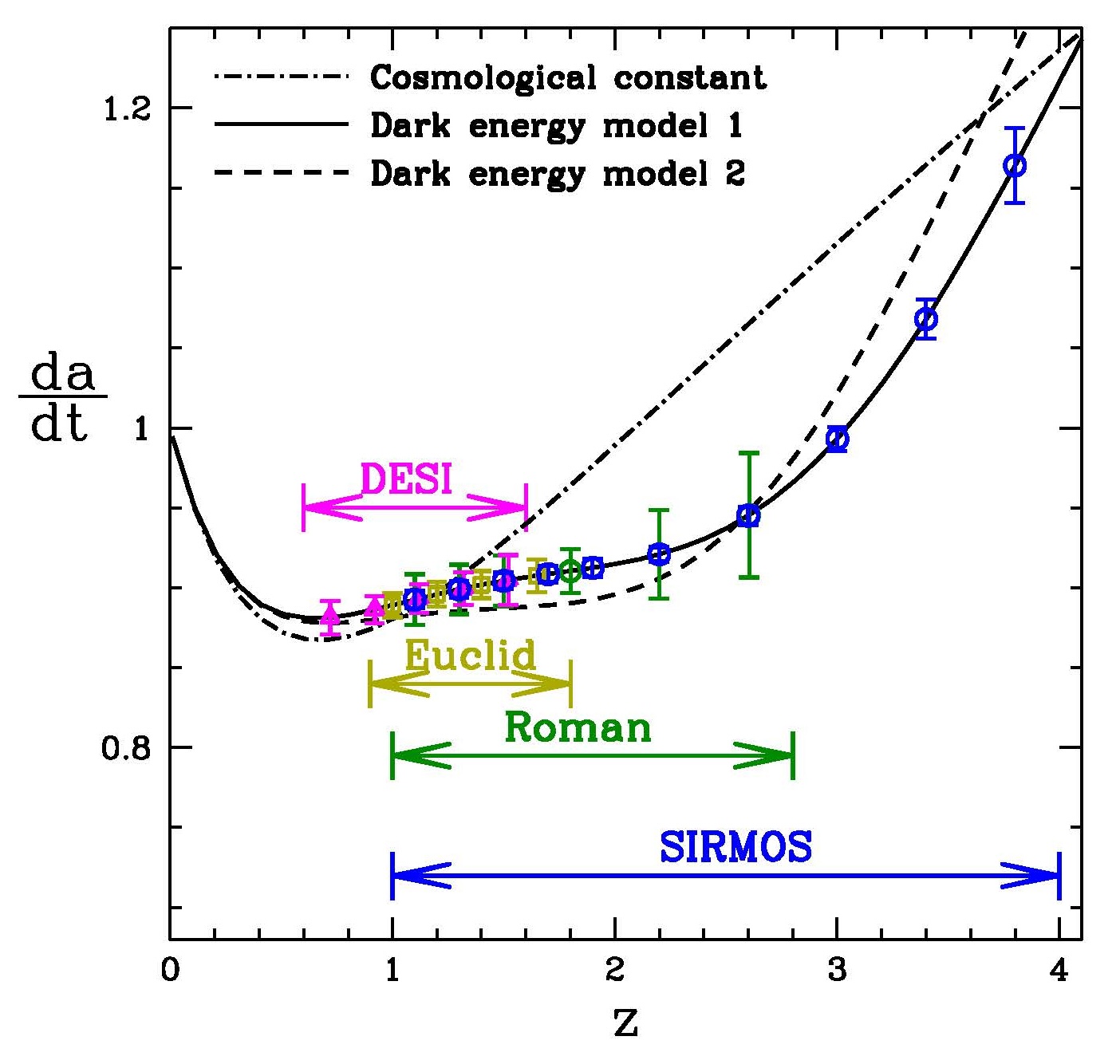} 
   \vspace{-0.1in}
   \caption{Left: Number density of galaxies detectable by different surveys. Centre: Forecast constraints on DE parameters in the CPL parametrisation from SIRMOS together with \emph{Planck} CMB in two scenarios motivated by the results in  \cite{Abdul_Karim_2025}, compared to equivalent DESI constraints. Right: Time derivative of cosmic scale factor vs. redshift (normalized to 1 at $z=0$), with predicted measurement uncertainties for various surveys. SIRMOS is able to distinguish DE models with more complex high-$z$ behaviour.
   }
   \label{fig:dadt}
   \vspace{-0.1in}
   \end{figure}
 
The fundamental nature of the dark energy (DE) driving the current epoch of accelerated expansion is unknown and remains the biggest open problem in cosmology. Recent results from DESI \cite{Abdul_Karim_2025} suggest a possible time evolution in the DE equation of state $w(z)$. If confirmed, this would rule out the cosmological constant model and constitute a major revolution in physics, necessitating a new fundamental field. The DESI results hint at a possible ``phantom" behaviour ($w(z)<-1$) at high redshifts, which raises serious theoretical challenges: models which accomplish this are hard to embed in well-behaved field theories without instabilities. However, current spectroscopic surveys, including DESI, have a notable gap in sensitivity in the redshift range $1.5\lesssim z\lesssim4$, coinciding with the inferred phantom behaviour. In addition, the DESI results, in common with many in the literature, use a phenomenological CPL parametrization of the potential time variation of the DE equation of state, $w(a) = w_0 + w_a(1 - a) $ \cite{Chevallier:2001,Linder2003}. This prescription however limits the space of models to those which have a constant equation of state equal to $w_0+w_a$ at high redshifts. Precise measurements of the expansion history in the high redshift gap are therefore essential to confirm or rule out DE evolution and to better investigate possible phantom dynamics. 
\\
\underline{\bf Objective 2:} \textbf{Measure the high-redshift expansion history with sufficient accuracy to distinguish models of late-time cosmic acceleration. }
A wide-area, high-throughput and highly multiplexed survey like SIRMOS, covering the redshift range $1<z<4$ with a target galaxy density far exceeding that of current and future facilities (Fig.~\ref{fig:dadt}), will be able to accurately measure both cosmic expansion and structure growth history
%distances 
using the baryon acoustic oscillation (BAO) feature \cite{Blake2003,Seo2003} as well as redshift-space distortion (RSD) features \cite{Kaiser1987} in the galaxy power spectrum. It will enable the expansion rate or time derivative of the scale factor to be determined at a precision far exceeding that from other planned surveys (Fig.2). 
The SIRMOS measurements can not only distinguish the CPL models from a cosmological constant (see Fig.2) -- without requiring the addition of data from type Ia supernovae, currently the most important source of potential systematic errors -- but crucially, they also directly test more general models of dark energy or modifications to General Relativity \cite{Guzzo2008,Wang2008}.

\vspace{-0.1in}
\subsection{Science Goal 3: Precisely measure the total neutrino mass sum}
Neutrinos are the only particle in the Standard Model of particle physics whose mass parameters remain unknown. The sum of the masses of all neutrino mass eigenstates, $\sum m_\nu$, can also 
%determine between 
be used to differentiate
the normal and inverted mass hierarchies. Terrestrial experiments set a lower bound on $\sum m_\nu>0.06$ eV \cite{2022PTEP.2022h3C01W}, but cosmological observations provide the best opportunity to unambiguously measure this mass scale. Current cosmological constraints require $\sum m_{\nu} < 0.064$ eV (at 95\% C.L.) \cite{Abdul_Karim_2025}, assuming a cosmological constant DE. In addition to the intrinsic importance of measuring the mass scale for the first time, the tension between the upper and lower bounds may itself be a sign of exotic physics scenarios \cite{Craig24}. However, forecasts for currently planned facilities only reach a precision of $\sigma(\sum m_\nu)$ in the range $0.015$ eV to $0.03$ eV \cite{2025JCAP...08..034A, 2025A&A...693A..58E}. Thus a 
%step‑change 
game changer in precision is required to obtain a high‑significance measurement of $\sum m_\nu$ 
if it is indeed close to the minimum allowed value.
\\
\underline{\bf Objective 3:} \textbf{Decisive measurement of the sum of neutrino masses.} Massive neutrinos suppress the growth of structure below the free‑streaming scale, imprinting a characteristic, scale‑ and redshift‑dependent signal in galaxy clustering. In the $1 < z < 4$ range that a 
SIRMOS-like mission would target fluctuations that are closer to linear, so the validity of perturbation theory models extends to smaller scales, and modelling systematics are easier to control. Our forecasts show that, in combination with \textit{Planck} CMB priors, a survey like SIRMOS will achieve a detection of the minimal mass scenario ($\sum m_\nu=0.057$~eV) at $>10\sigma$ significance through combining the power spectrum, bispectrum and multi-tracer analysis techniques, decisively establishing the absolute neutrino mass scale. 
%%% end of Sesh replacement text %%%%

\vspace{-0.1in}
\section{Proposed approach, mission concept and feasibility}
\underline{\bf Proposed Approach:} Meeting the SIRMOS science objectives requires a galaxy redshift survey capable of mapping the 3D large-scale structure of the Universe with high fidelity across a very large cosmological volume. The key scientific drivers demand access to ultra-large scales, dense samples of tracers, and exceptional control of systematic uncertainties. Existing and forthcoming missions do not provide the combination of depth, number density, and spectroscopic precision needed to fully meet these requirements; see Fig. \ref{fig:dadt}. 

\begin{wrapfigure}{r}{0.4\linewidth}
\vspace{-0.2in}
%begin{figure}
    \centering
    \includegraphics[width=\linewidth]{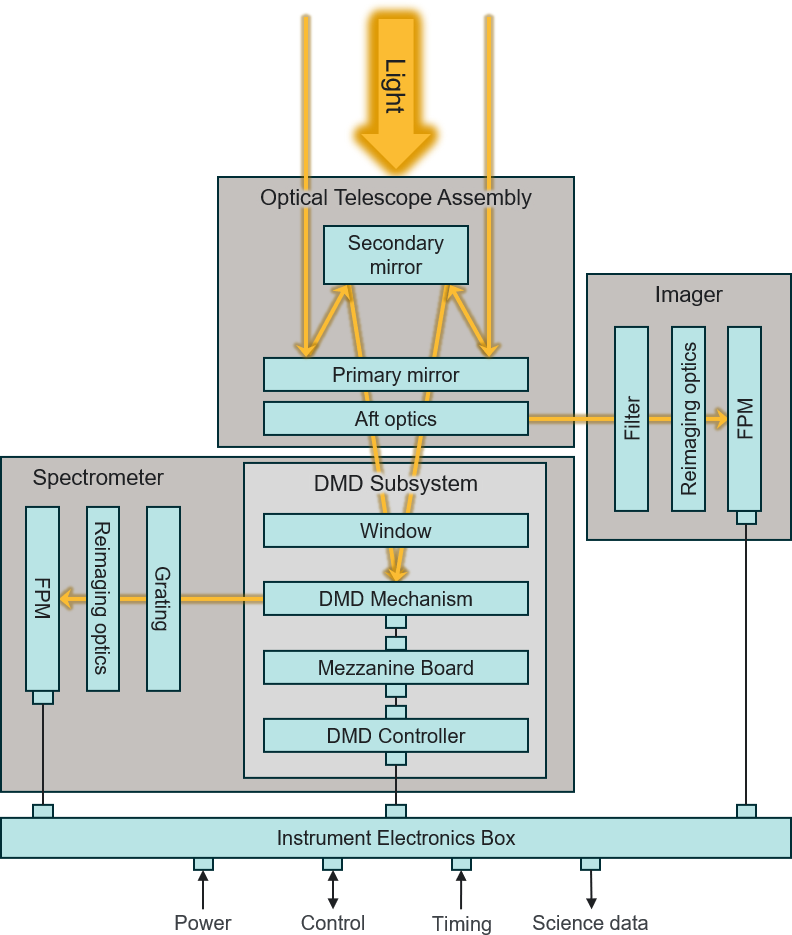} %\hspace{0.4in}
    \vspace{-0.2in}
    \caption{SIRMOS payload Functional Block Diagram.}
    \vspace{-0.15in}
    \label{fig:fbd}
%\end{figure}
\end{wrapfigure}

SIRMOS science objectives require observing conditions in which the full spectral range needed for precise redshift determination is continuously accessible, where backgrounds are stable, and where calibration can be maintained uniformly over the lifetime of the survey. From the ground, atmospheric absorption, variable transmission, and dense variable OH sky emission lines fragment the NIR spectral window, complicating line identification and reducing completeness. Beyond $1.1~\mu$m, low atmospheric transmission and rapidly varying backgrounds introduce systematic uncertainties that are difficult to model, particularly for ultra-large-scale measurements. A space-based platform naturally overcomes these limitations by providing continuous wavelength coverage from $1.1$ to $2.5~\mu$m, a stable low background, and uniform calibration conditions, enabling systematic uncertainties to be controlled at the level required for unbiased measurements of PNG, dark energy, and neutrino mass.

To meet the science requirements, SIRMOS will conduct a galaxy redshift survey to trace the large-scale structure of the Universe in unprecedented detail, covering $14{,}000~\mathrm{deg}^2$ over $1 < z < 4$, corresponding to a cosmic volume of $\sim 500~\mathrm{Gpc}^3$.
The survey will obtain spectroscopic redshifts for more than $10^8$ galaxies identified through H$\alpha$, [O\,\textsc{iii}], and [O\,\textsc{ii}] emission lines. This dense and homogeneous map of cosmic structure will enable precision measurements of fundamental physics, while maintaining tight control of systematic errors. SIRMOS redshift accuracy of $\sigma_z / (1+z) \simeq 10^{-4}$, achieved through slit spectroscopy, represents an order of magnitude improvement over Euclid and Roman ($\sim$$10^{-3}$ via slitless grisms) and several orders beyond SPHEREx ($\sim$$10^{-1}$ for the cosmology sample, using linear variable filters). 
SIRMOS leverages technology from previous successful work by BAE Space \& Mission Systems (BAE-SMS) on other missions.

\noindent
\underline{\bf Mission Concept:} SIRMOS will be a NASA MIDEX mission in a sun-synchronous orbit, with a SPHEREx-like spacecraft bus and payload thermal design, and a 60~cm diameter telescope with $\sim$ 1 deg$^2$ FoV, using a Digital Micro-mirror Device (DMD) to obtain $\sim$ 4000 slit spectra simultaneously. To obtain $10^8$ galaxy spectra at $1<z<4$, the exposure time per field is 51~min, acquired in 3 separate 17-min visits made at 12-month intervals during the 3-year prime mission. After each visit, galaxies with spectra at $S/N \geq 5$ will be replaced with new galaxy targets.
The spacecraft will keep the payload inertially fixed for the 17-minute integration time (split into multiple exposures for dithering and cosmic ray rejection), then slew to the next pointing.

\begin{table}[h!]
    % \vspace{-0.1in}
    \centering
    \includegraphics[width=0.55\linewidth]{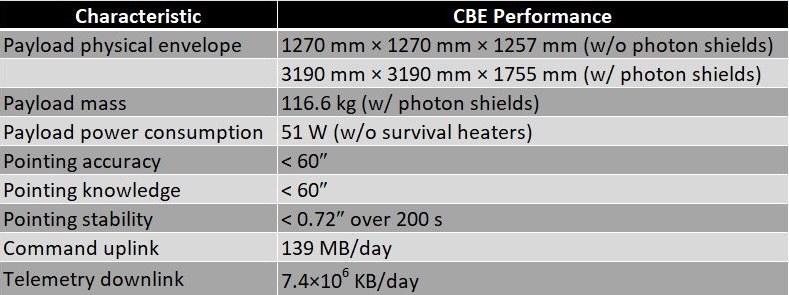}\hspace{0.05in}
    \includegraphics[width=0.42\linewidth]{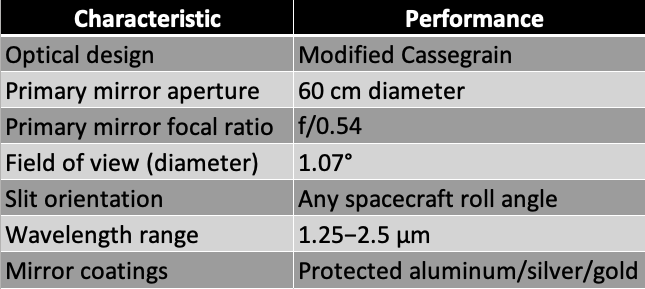}
    % \vspace{-0.1in}
    \caption{Left: SIRMOS payload key parameters.
    Right: SIRMOS telescope characteristics.}
    \label{tab:sirmosParameters}
    % \vspace{-0.1in}
\end{table}

\begin{figure}
    \centering
    \includegraphics[width=0.32\linewidth]{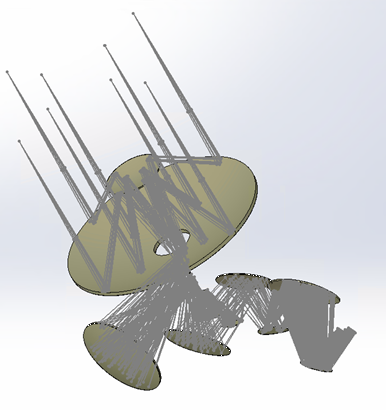}\hspace{0.2in}
    \includegraphics[width=0.53\linewidth]{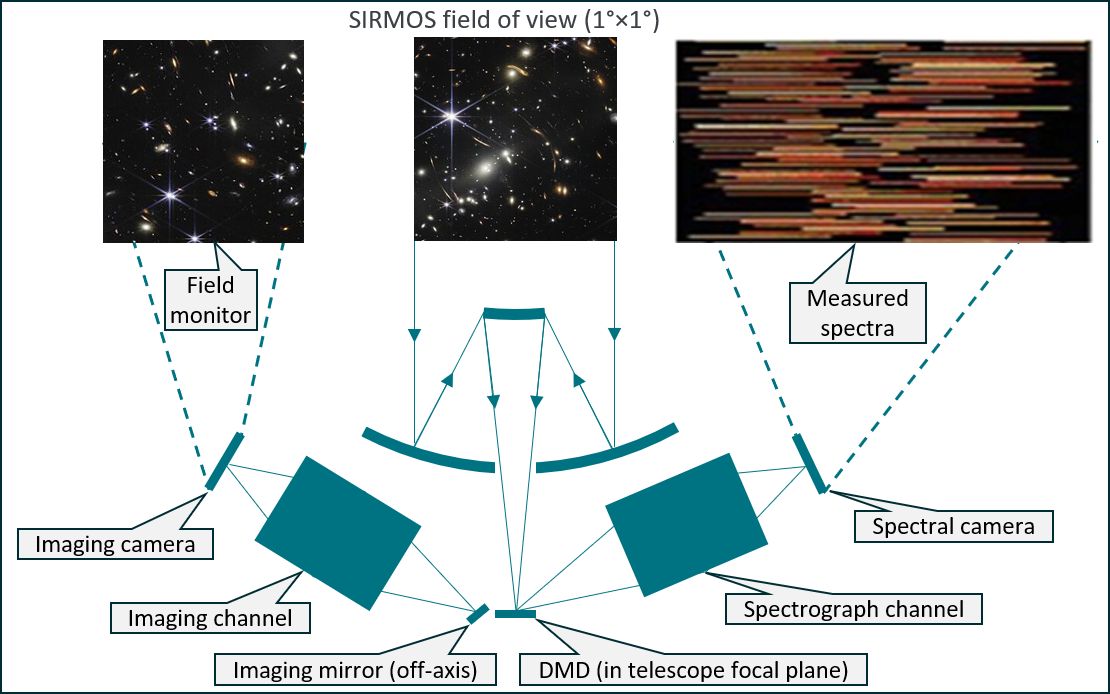}
    % \vspace{-0.1in}
    \caption{Left: SIRMOS telescope elements deliver wide-field, diffraction-limited images to the DMD.
    Right: DMD directs light from 4000 selected targets to the spectral channel for highly-mutiplexed operation. }
    \label{fig:aperture}
    % \vspace{-0.2in}
\end{figure}

\noindent
\underline{\bf Payload Overview:} To achieve its ambitious goal of collecting $10^8$ redshifts over the 3-year prime mission lifetime, the SIRMOS payload must deliver a wide FoV of $\sim$ 1 deg$^2$ with exceptionally flexible multi-object spectroscopy capability (4000 spectra simultaneously) in a small package. As shown in Fig.~\ref{fig:fbd}, the payload achieves this with a simple design incorporating an imager and spectrometer within a compact volume featuring only one mechanism. The telescope and aft optics feed light to a DMD, which activates selected micro-mirrors to redirect light for specific targets to the spectrometer for redshift measurements. An off-axis imager views a field adjacent to the spectroscopy field to monitor on-sky alignment. Both the spectrometer and imager feature low-noise array detectors for high S/N observations. Table~\ref{tab:sirmosParameters} (left) lists the key payload parameters.

\noindent
\underline{\bf Telescope Requirements \& Estimates:} SIRMOS requires a wide FoV coupled with sufficient light-gathering power to observe key emission lines.
The main telescope parameters are listed in Table~
\ref{tab:sirmosParameters} (right).
The telescope aperture diameter of 60~cm meets the science requirements with margin. 
Payload cost is not strongly coupled to the aperture size as long as the total payload mass remains within the limit imposed by the bus.
Fig.~\ref{fig:aperture} (left)
shows a ray trace of the all-reflective design which is cooled to 140~K to limit thermal background. The all-aluminium telescope design features numerous aspheric surfaces, which preserve image quality; future iterations of the optical design will consider ease of manufacturability and integration as key constraints. 
The pointing requirements for SIRMOS in Table~\ref{tab:sirmosParameters} (left) are driven by the science requirements, and are achievable using conventional technology BAE-SMS has already deployed on other missions. 

\begin{wraptable}{r}{0.48\linewidth}
\includegraphics[width=\linewidth]{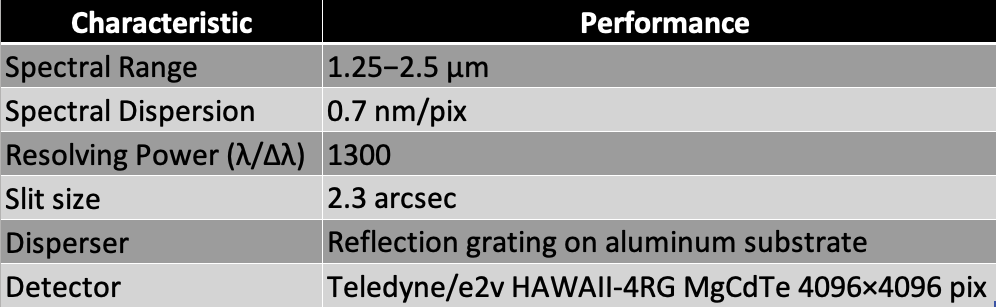}
    \vspace{-0.32in}
    \caption{SIRMOS spectrometer parameters.}    
    \label{tab:spectrometerParameters}
    \vspace{-0.12in}
\end{wraptable}

\noindent
\underline{\bf Spectrometer Requirements and Estimates:} The SIRMOS spectrometer is designed to be both powerful and simple, leveraging the heritage from the SAMOS DMD-based spectrograph developed by SIRMOS team members and successfully deployed on the SOAR telescope in Chile \cite{10.1117/12.2311876,10.1117/12.3020796}. 
Table~\ref{tab:spectrometerParameters} lists spectrometer characteristics.
Figure~\ref{fig:aperture} (right) illustrates how the SIRMOS payload employs the DMD for spectroscopy.
The DMD is at an appropriate level of technological readiness for the conceptual stage of SIRMOS. 
DMDs have been successfully tested for robustness and performance in space by ground-based tests \cite{2017arXiv170806241T} and
a DMD will be flown on a NASA-sponsored rocket experiment in 2026 to establish compatibility with the space environment \cite{10.1117/12.2594651,2025AAS...24614105P}. The remaining task is to demonstrate operation in the NIR; DMDs have been successfully tested at cryogenic temperatures, but their native 
window on the hermetically-sealed unit is opaque beyond 1.0~$\mu$m. The SIRMOS team is developing strategies to 
replace the visible window with one transparent to NIR radiation,
a straightforward engineering study presenting low technical risk.
SIRMOS captures spectra with a Teledyne HAWAII-4RG
(H4RG) 4K$\times$4K HgCdTe focal plane array (FPA) \cite{10.1117/12.926750}, with space heritage on the Roman Space Telescope Wide Field Imager (WFI),
and well suited to the mission by virtue of its efficient power and noise characteristics. 
The resulting spectrometer collects 4000 spectra simultaneously,
enabling the collection of $10^8$ redshifts over the 3-year prime mission.

\begin{wrapfigure}{R}{0.55\textwidth}
% \vspace{-0.2in}
%\begin{figure}
    \centering
    \includegraphics[width=\linewidth]{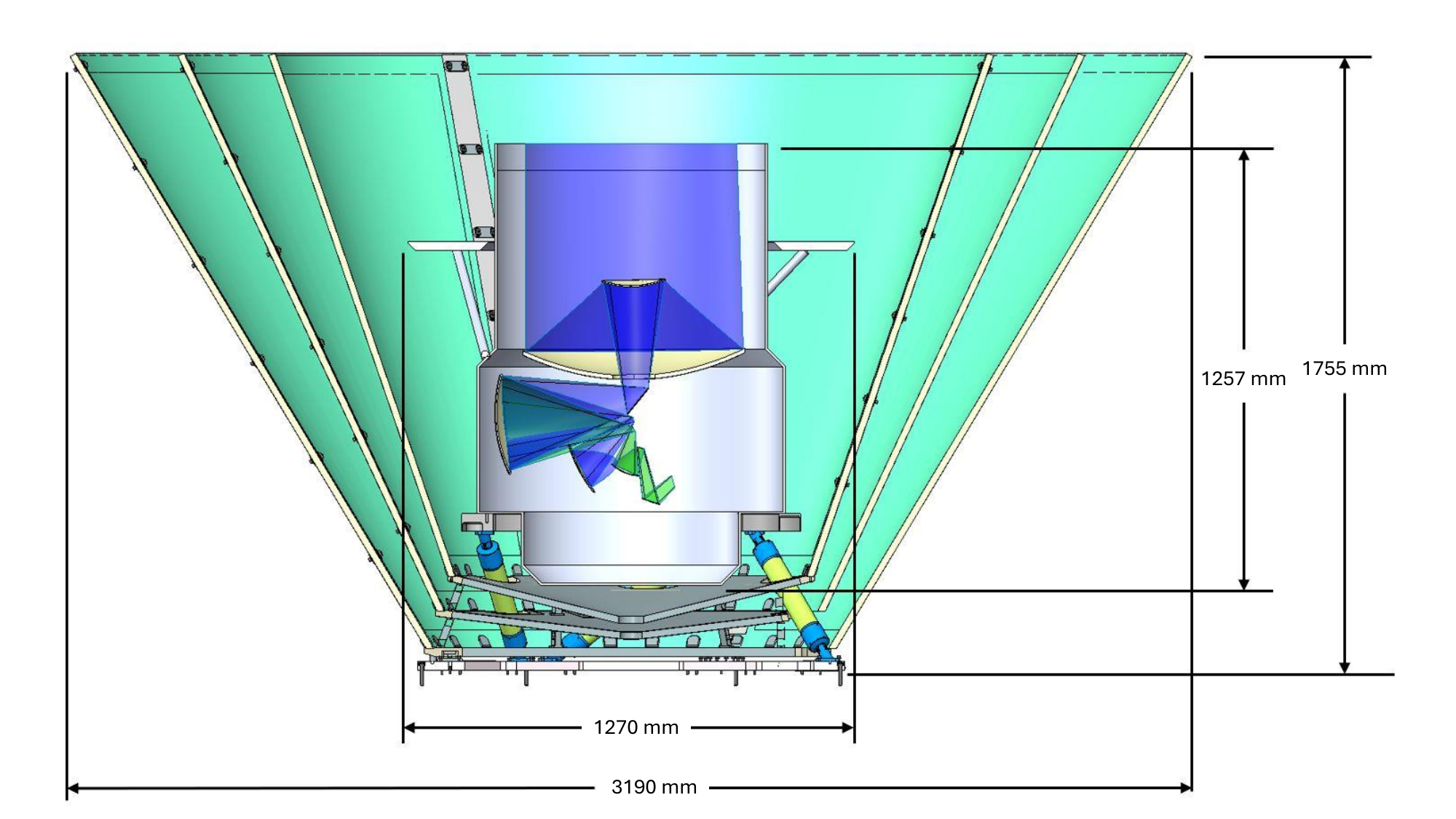}
    \vspace{-0.3in}
    \caption{The SIRMOS payload is well-sized to fit within the SPHEREx photon shields.}
    \vspace{-0.08in}
    \label{fig:sirmosVolume}
%\end{figure}
\end{wrapfigure}

\noindent
\underline{\bf Imager Requirements and Estimates:} The SIRMOS imager's primary function is to sense the location of pre-selected alignment stars used to confirm the telescope pointing and sky position angle, thereby ensuring that the defined DMD slit pattern for a given field aligns with the sky positions of the spectroscopic targets. The imager is fed from an off-axis pickoff mirror beside the DMD and thus views a region of sky which is close to but not overlapping with the spectrometer FoV. The imager's wavelength coverage is dictated by the wavelength response of the detector, which can be either visible or NIR. The imager requires sufficient FoV and spatial resolution to measure the pointing to better than $0\farcsec1$ and the sky position angle to 10\arcsec. These requirements dictate a minimum FoV of 6\arcmin\ and spatial sampling of 0\farcsec1. 

\begin{wraptable}{r}{0.45\textwidth}
\vspace{-0.15in}
\centering
  \includegraphics[width=\linewidth]{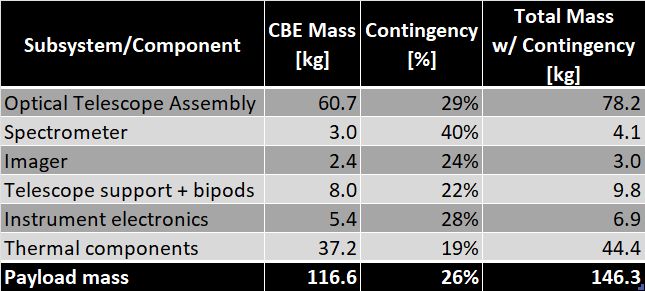}
\vspace{-0.2in}
\caption{SIRMOS payload mass properties.}
% \vspace{-0.2in}
\label{tab:sirmosMass}
\end{wraptable}

\noindent
\underline{\bf Thermal Performance:} SIRMOS leverages the successful thermal design of SPHEREx.
The fully passive thermal design uses three V-groove radiators \cite{Moore_2024}. 
The SIRMOS thermal design easily meets the payload's needs with plenty of margin.

\noindent
\underline{\bf Mass and Dimensions:} 
Table~\ref{tab:sirmosMass} shows a breakdown of the CBE mass by subsystem. 
The SIRMOS payload volume is well sized to fit within the SPHEREx radiators, see Fig.~\ref{fig:sirmosVolume}, which 
gives the dimensions of both the inner portion of the payload (excluding the V-groove radiators, light cones, and bipods) and the full payload, including those thermal and structural elements. The SPHEREx bus can easily accommodate the SIRMOS payload mass and volume.

\noindent
\underline{\bf Power Requirements:}
SIRMOS payload has relatively low power needs (Table~\ref{tab:sirmosPower}), well within the SPHEREx bus capabilities.

\begin{wraptable}{r}{0.5\textwidth}
% \vspace{-0.2in}
\centering
\includegraphics[width=\linewidth]{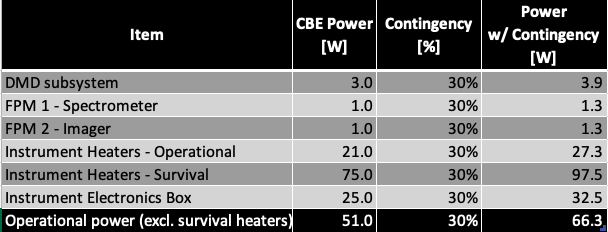}
% \vspace{-0.3in}
\caption{SIRMOS payload power requirements.}
\vspace{-0.1in}
\label{tab:sirmosPower}
\end{wraptable}

\noindent
\underline{\bf Telemetry Requirements:} 
The total daily SIRMOS data volume transmitted to the ground is 7.4~GB,
well within the 97~GB daily data volume 
SPHEREx transmits daily using its Ka-band antenna.
A week-long observing plan amounts to only 139~MB %(139,000~kB) 
per day, easily handled by a 2,000~kbps S-band receiver. SIRMOS data needs are well within the spacecraft capacity and a SPHEREx-like space-to-ground communication system.

\noindent
\underline{\bf Technology Readiness and Heritage:} The SIRMOS payload design leverages heavily the successful design heritage of SPHEREx. The DMD is currently at NASA Technology Readiness Level (TRL) 5 (pending). A planned rocket mission equipped with a DMD will fly in 2026 to advance the TRL to 6. All remaining payload items are at $\textrm{TRL}\ge6$.
\vspace{-0.15in}

%{-0.1in}
\section{Global and strategic context: UK leadership and partnership opportunities}
\vspace{-0.1in}

SIRMOS is a proposed NASA MIDEX mission, to which the UK can make important instrumentation contributions in partnership with NASA. SIRMOS offers a unique opportunity for UK scientists to join and lead groundbreaking cosmology research, as well as for UK industrial partners to contribute cutting-edge instrumentation, e.g. complex telescope mirrors, advanced detector applications, data processing pipelines, and autonomous target acquisition. The UK contribution to SIRMOS would focus on building the telescope and its optical systems, drawing on expertise in high-precision fabrication and integration.
SIRMOS also builds directly on the UK’s substantial investment in ESA's Euclid mission: UKSA has funded the instrument and infrastructure development for Euclid for over a decade, while STFC has supported UK teams leading Euclid science analyses. SIRMOS leverages this expertise, using Euclid photometry for target selection. It represents a natural continuation of the UK's global leadership in large-scale structure cosmology and analysis of spectroscopic surveys, including both Euclid and the ground-based DESI. By extending the redshift reach in a focused NASA/UK partnership, SIRMOS offers extraordinary scientific capabilities with a smaller, agile mission architecture.

SIRMOS addresses frontier science priorities identified in the STFC Astronomy Advisory Panel 2022 Roadmap \cite{2023arXiv230105457S}. The roadmap notes science themes to be addressed in the next two decades include the nature and composition of dark energy. It further highlights that UK astronomy spans “the evolution of galaxy populations and the large scale structure of the Universe, the properties of dark matter and dark energy, [and] the fundamental cosmological parameters and theories of gravity.” By delivering world-class measurements of galaxy clustering and cosmic expansion at redshifts 1 to 4, SIRMOS aligns STFC’s scientific leadership with UKSA priorities and engages UK industry in high-value manufacturing. It also strengthens domestic capability in flight-qualified instrumentation, linking fundamental physics to national goals including high-technology exports, skilled employment, and end-to-end mission delivery from UK facilities.

In addition to the science case for inflation, dark energy and neutrinos outlined above, a high-$z$ NIR survey such as SIRMOS would also make enormous contributions to other complementary areas, such as determining how environment shapes galaxy evolution at high redshifts and probing the Milky Way's dust-enshrouded regions. This enables strong synergies with existing UK-led ground-based capabilities in the NIR, such as MOONRISE \cite{2020Msngr.180...24M} operating on MOONS (Multi-Object Optical and Near-IR Spectrograph), which is managed by STFC and UKATC. Leveraging existing UK leadership with new opportunites at the frontiers of cosmology make this a very attractive mission proposal.

\newpage
\bibliographystyle{IEEEtranDOI} % sort in order of appearance
\bibliography{sirmosReferences} % see instructions for creating references in file AAAREADME.txt in this overleaf folder
\end{document}